\documentclass[12pt]{article}
\newif\ifpdf\ifx\pdfoutput\undefined\pdffalse\else\pdfoutput=1\pdftrue\fi
\newcommand{\pdfgraphics}{\ifpdf\DeclareGraphicsExtensions{.pdf,.jpg}
\else\fi} 
\usepackage{graphicx,epstopdf,amsfonts,amsmath,array,mathrsfs}
\newcommand{\pbs}[1]{\let\temp=\\#1\let\\=\temp}
\numberwithin{equation}{section}
\def\be{\begin{equation}}\def\ee{\end{equation}}
\def\cvp{\raise 2pt\hbox{,}}  \def\cvd{\raise 2pt\hbox{.}}

 \def\Tr{\mathop{\rm
Tr}\nolimits}

\def\d{{\rm d}}
\def\nn{{\cal N}}

\def\Nf{N_{\text f}}
 \def\uN{{\rm U}(N)} 
  
\def\La{\Lambda}
\def\mic{\text{mic}}

\def\wmic{W_{\text{mic}}}

\def\a{\boldsymbol{a}}\def\b{\boldsymbol{b}}
\def\g{\boldsymbol{g}} 
 \def\m{\boldsymbol{m}}\def\t{\boldsymbol{t}}

\def\u{\text{U}(1)}
 \def\eps{\epsilon}
\def\ranglecl{\rangle_{\text{cl}}}
\def\vevas#1{\langle\a|#1|\a\rangle}
\def\vevab#1{\bigl\langle\a\big|#1\big|\a\bigr\rangle}

\def\cpart{\vec{\mathsf k}} \def\k{\mathsf k} \def\0{\vec{\mathsf{0}}}

\def\npb#1#2#3{{\it Nucl.\ Phys.\ }{\bf B #1} (#2) #3}

\def\jhep#1#2#3{{\it JHEP\ }{\bf #1} (#2) #3}
\def\prd#1#2#3{{\it Phys.\ Rev.\ }{\bf D #1} (#2) #3}

\def\atmp#1#2#3{{\it Adv.\ Theor.\ Math.\ Phys.\ }{\bf #1} (#2) #3}

\begin{document}
\pdfgraphics 
%
\pagestyle{empty}
\parskip 0in

\hfill LPTENS-09/06


\vfill
\begin{center}
{\LARGE Flavors in the microscopic approach to}

\bigskip

{\LARGE $\mathcal N=1$ gauge theories}

\vspace{0.4in}

Frank \textsc{Ferrari} and Vincent \textsc{Wens}
\\
\medskip
{\it Service de Physique Th\'eorique et Math\'ematique\\
Universit\'e Libre de Bruxelles and International Solvay Institutes\\
Campus de la Plaine, CP 231, B-1050 Bruxelles, Belgique
}\\
\smallskip
{\tt fferrari@ulb.ac.be, vwens@ulb.ac.be}
\end{center}
\vfill\noindent

In this note, we solve an extended version of the $\nn=1$ super
Yang-Mills theory with gauge group $\uN$, an adjoint chiral multiplet
and $\Nf$ flavors of quarks, by using the $\nn=1$ microscopic
formalism based on Nekrasov's sums over colored partitions. Our main
new result is the computation of the general mesonic operators. We
prove that the generalized Konishi anomaly equations with flavors are
satisfied at the non-perturbative level. This yields in particular a
microscopic, first principle derivation of the matrix model disk
diagram contributions that must be included in the Dijkgraaf-Vafa
approach.

\vfill

\medskip
%
\begin{flushleft}
\today
\end{flushleft}
\newpage\pagestyle{plain}
\baselineskip 16pt
\setcounter{footnote}{0}

\section{Introduction}
\setcounter{equation}{0}

The recently developed microscopic formalism to $\nn=1$ supersymmetric
gauge theory \cite{mic1, mic2, mic3} is a first-principle approach
that allows in principle to solve rigorously a general $\nn=1$ gauge
theory in the chiral sector. It is based on the direct calculation of
the relevant path integrals and relies heavily on Nekrasov's instanton
technology \cite{nekrasov}, suitably adapted to the $\nn=1$ context.
The goal of the present paper is to develop the formalism in the case
where fundamental quark flavors are present. Including quarks brings
several new interesting features that we shall explain in details in
the following.

The model we consider is the $\uN$ supersymmetric gauge theory with an
adjoint chiral multiplet $X$ and chiral multiplets $Q_{f}$ and $\tilde
Q^{f}$, $1\leq f\leq\Nf$, in the fundamental and anti-fundamental
representations respectively. We shall restrict ourselves to the cases
$\Nf\leq 2N$.

The lagrangian is given by
\begin{multline}\label{action} \mathcal L =
\mathrm{Re}  \int \d^2 \theta\, \frac{\tau}{4\pi i} \Tr
W^{\alpha}W_{\alpha} + \int \d^2\theta \d^2\bar{\theta}\, \Tr
X^{\dagger} e^{2V} X \\ +  \int \d^2\theta \d^2\bar{\theta}\,
Q^{\dagger f} e^{2V} Q_f  +  \int \d^2\theta \d^2\bar{\theta}\,
\tilde{Q}^{\dagger}_{\tilde f}\, e^{2V} \tilde{Q}^{\tilde f} 
+ 2N\, \mathrm{Re} \int \d^2\theta \ \mathscr W \, ,
\end{multline}
where
\be\label{taudef}\tau=\frac{\vartheta}{2\pi} + i\frac{4\pi}{g^{2}}\ee
is the gauge coupling constant and
\begin{equation}\label{W}
\mathscr W = \frac{-1}{16\pi^2} \Tr \big( V(X)W^{\alpha}W_{\alpha}
\big) + \Tr W(X) + {}^T\tilde Q^{\tilde f} m_{\ \tilde f}^{f}(X) Q_f\,
.\ee
The polynomials $V(z)$, $W(z)$ and $m^{f}_{\ \tilde f}(z)$ are
parametrized as follows:
\begin{align} \label{Vz} &NV(z) = \sum_{k \geq 2} 
k \, t_k\, z^{k-1} \, , \\ \label{Wz} &W(z) =  \sum_{k \geq 0} 
\frac{g_k}{k+1}\, z^{k+1} \, , \\ 
\label{mz} &m_{\ \tilde f}^{f}(z) = 
\sum_{k \geq 0} m_{k, \tilde f}^{f} \, z^k \, . \end{align}
It is useful to introduce
\be \label{Uz} U(z) = \det \big( m(z) \big) = U_0 \prod_{Q= 1}^{\ell}
(z-b_Q) \,  \ee
and 
\be \label{tz} t(z) = \sum_{k \geq 1} \frac{t_k}{k+1}\, z^{k+1} \,
,\ee 
with
\be\label{defq} t_1 = \ln (U_0 q)\, .  \ee
The coupling $q$ is the instanton factor in the model. In the
asymptotically free case ($\Nf<2N$) it is expressed in terms of the
dynamically generated scale $\La$ as
\be\label{q1} q = \La^{2N-\Nf}\, ,\ee
whereas in the case of vanishing $\beta$ function ($\Nf=2N$) we have
\be\label{q2} q = e^{2\pi i \tau} = e^{i\vartheta-8\pi^{2}/g^{2}}\,
.\ee
The set of couplings $\{t_k\}_{k\geq2}$, $\{g_k\}$,
$\{m^{f}_{k,\,\tilde f}\}$ and $\{b_Q\}$ will be denoted by $\t$,
$\g$, $\m$ and $\b$ respectively. In the most standard case, $t_{k}=0$
for $k\geq 2$, but as explained in \cite{mic3} it is actually very
natural and convenient to consider the more general theory with
arbitrary couplings $\t$.

The chiral ring of the model is generated by the operators 
\begin{equation}\label{chirops} \Tr X^k \, , \quad \Tr
W^{\alpha}W_{\alpha} X^k \, , \quad {}^T\tilde Q^{\tilde f} X^k Q_f
\end{equation}
and our goal is to compute, from first principles, the corresponding
expectation values. These expectation values are multi-valued analytic
functions of the parameters $\t$, $\g$, $\m$ and $q$. The
multi-valuedness comes from the existence of several distinct vacua in
the theory. At the classical level, the most general supersymmetric
vacuum, which is obtained by extremizing the tree-level
superpotential, can be labeled as $|N_{i};\nu_{Q}\ranglecl$, where the
$N_{i}\geq 0$ and $\nu_{Q}\in \{0,1\}$ are integers satisfying the
constraint
\be\label{NX}\sum_{i=1}^{\mathrm{deg}\, W'}N_{i} + \sum_{Q=1}^{\ell}
\nu_{Q} = N\, .\ee
The $N_i$s and $\nu_Q$s denote the number of eigenvalues of the
matrix $X$ that are equal classically to the $i^{\textrm{th}}$ root of
$W'(z)$ and to $b_Q$, respectively. The $\uN$ gauge symmetry in a
vacuum $|N_{i};\nu_{Q}\ranglecl$ is thus broken down to a product of
$\text{U}(N_{i})$ factors. The number of non-trivial factors of the
unbroken gauge group, i.e.\ the number of non-zero $N_{i}$s, is called
the rank of the vacuum. In the quantum theory, chiral symmetry
breaking yields a larger degeneracy of the vacua, that are then
labeled as $|N_{i},k_{i};\nu_{Q}\rangle$ where $0\leq k_{i}\leq
N_{i}-1$. This structure follows from the extremization of the
microscopic quantum superpotential that will be introduced in the next
Section.

The expectation values of the operators \eqref{chirops} are most
conveniently encoded in the generating functions
\begin{align} \label{Rdef} \mathscr R(z) &= \sum_{k\geq 0}
\frac{\langle \Tr X^k \rangle} {z^{k+1}}\, \cvp\\ \label{Sdef}\mathscr
S(z) & = -\frac{1}{16\pi^2}\sum_{k\geq 0} \frac{\langle \Tr
W^{\alpha}W_{\alpha} X^k \rangle}{z^{k+1}}\,\cvp\\
\label{Gdef}\mathscr
G^{\tilde f}_{\ f}(z)& = \sum_{k\geq 0}\frac{\langle {}^T\tilde
Q^{\tilde f} X^k Q_f \rangle}{z^{k+1}}\, \cdotp
\end{align}
One of our main result is to show that these generating functions
satisfy the following set of algebraic equations,
\begin{gather} 
\label{a1} NW'(z) \mathscr S(z) - \mathscr S(z)^{2} = \Delta_{\mathscr
S} (z)\, , \\
\label{a2} N m_{\ \tilde f}^{f'}(z)\mathscr G^{\tilde f}_{\ f}(z) -
\mathscr S(z)\delta_{f}^{f'} = \Delta_{f}^{f'}(z)\, , \\
\label{a3} N\mathscr G^{\tilde f'}_{\ f}(z)m_{\ \tilde f}^{f}(z) -
\mathscr S(z)\delta_{\tilde f}^{\tilde f'} =\tilde\Delta_{\tilde
f}^{\tilde f'}(z)\, , \\
\label{a4} t'''(z)\mathscr S(z)+NW'(z)\mathscr R(z) + N{m'}_{\ \tilde
f}^{f}(z) \mathscr G^{\tilde f}_{\ f}(z) - 2 \mathscr S(z)\mathscr
R(z) = \Delta_{\mathscr R}(z)\, ,
\end{gather}
where $\Delta_{\mathscr S}$, $\Delta_{f}^{f'}$, $\tilde\Delta_{\tilde
f}^ {\tilde f'}$ and $\Delta_{\mathscr R}$ are polynomials. These
equations are the famous generalized Konishi anomaly equations
\cite{CDSW}, adapted to the case where flavors are included in the
model \cite{seifla} and suitably generalized to the extended theory
corresponding to having arbitrary couplings $\t$. These equations are
at the heart of the Dijkgraaf-Vafa matrix model formalism 
\cite{DV, argfla}, where they follow directly from the planar loop
equations of the matrix model. 
They were understood at the perturbative level (i.e.\ in
a fixed classical background gauge field) in \cite{CDSW,seifla}, but a
full non-perturbative proof requires much more work as explained in
great details in \cite{mic2,mic3}. This is where the microscopic
formalism shows its full power. It is remarkable to reproduce the
planar matrix model result (which, when flavors are present, also
include disk diagrams \cite{argfla}) from \emph{finite} $N$ gauge 
theory path integral calculations. As we shall see, these integrals can 
be reduced to non-trivial sums over colored partitions.

The constraints \eqref{a1}, \eqref{a2}, \eqref{a3} and \eqref{a4} do
not fix completely the expectation values. There remains undetermined
coefficients in the polynomials that appear in the right hand side of
these equations. This ambiguity is completely removed by the fact that
the number of colors $N$ in the gauge theory is finite and thus only a
finite number of the operators \eqref{chirops} are algebraically
independent \cite{per,ferwens,ferphases}. Mathematically, this is
translated into quantization conditions for the periods of $\mathscr
R\,\d z$,
\be\label{qc} \oint \mathscr R(z) \d z \in 2\pi i \mathbb Z\, .\ee
In our microscopic formalism, this is satisfied by construction, since
the operators are built explicitly from finite $N\times N$ matrices
and the relations \eqref{qc} will be easy to check.\footnote{In the
Dijkgraaf-Vafa matrix model formalism, on the other hand, the
conditions follow from the extremization of the postulated glueball
superpotential with all the required Veneziano-Yankielowicz terms
included \cite{CSWphases} and are thus highly non-trivial.}

The plan of the paper is as follows. In Section 2, we explain the
microscopic formalism in the case of the model \eqref{action}. In
Sections 3, 4 and 5 we compute the scalar, glueball and meson
operators respectively. Finally, in Section 6, we extremize the
microscopic superpotential, derive the anomaly equations and discuss
some general features of the solution. We have also included an
Appendix containing some technicalities used in the main text.

\section{The microscopic formalism}
\setcounter{equation}{0}

The starting point of the microscopic formalism \cite{mic1} is to
consider the expectation values of the operators \eqref{chirops}
\emph{with fixed boundary conditions at infinity} for the adjoint
scalar field $X$,
\be \label{bc} X_{\infty} = \text{diag} \ (a_1,\ldots, a_N) \, .\ee
The eigenvalues at infinity
\be\label{defa} \a = (a_1,\ldots, a_N)\ee
are arbitrary fixed complex numbers. We could also try to impose
arbitrary boundary conditions at infinity for the quark fields $Q_{f}$
and $\tilde Q^{\tilde f}$, but we prefer in the present paper to first
integrate over these fields \emph{exactly} in the path integrals. The
model is then reduced to the case with no flavor, but with extra
determinant-like factors due to the integration over the quarks. The
expectation value of an arbitrary chiral operator $\mathscr O$ with
the boundary conditions \eqref{bc} will be denoted by $\langle \a |
\mathscr O |\a \rangle$ and the corresponding generating functions in
our model are given by
\begin{align} \label{Rz} R(z; \a) &= \sum_{k\geq 0} 
\frac{\vevas{\Tr X^k}}{z^{k+1}}\, \cvp \\ \label{Sz} S(z;\a)& =
-\frac{1}{16\pi^2}\sum_{k\geq 0} \frac{\vevas{\Tr W^{\alpha}W_{\alpha}
X^k}}{z^{k+1}}\, \cvp \\ \label{Gz} G^{\tilde f}_{\ f}(z;\a) &=
\sum_{k\geq 0}\frac{\vevas{{}^T\tilde Q^{\tilde f} X^k Q_f
}}{z^{k+1}}\, \cvd
\end{align}
Clearly, the expectation values $\vevas{\mathscr O}$ are not physical
and the corresponding generating functions \eqref{Rz}, \eqref{Sz} and
\eqref{Gz} do not coincide with the physical generating functions
\eqref{Rdef}, \eqref{Sdef} and \eqref{Gdef}.\footnote{We have changed
slightly our notation with respect to references
\cite{mic1,mic2,mic3}; $R$ and $S$ in the present paper correspond to
$R_{\mic}$ and $S_{\mic}$ in our previous works.} In particular, the
functions $R$, $S$ and $G^{\tilde f}_{\ f}$ depend on the arbitrary
boundary conditions $\a$ (as well of course on the other parameters in
the model), whereas the physical correlators depend on a choice of
vacuum but not on $\a$.

The interest in considering the correlators for fixed boundary
conditions at infinity is that, at least in an open set in $\a$-space,
they can always be computed by summing a convergent instanton series
\cite{mic1}. The functions at arbitrary $\a$ (outside the radius of
convergence of the instanton series) are then obtained uniquely by
analytic continuations.

There exists a quantum superpotential $\wmic(\a)$ for the boundary
conditions $\a$ \cite{mic1}. One of the fundamental property of this
quantum superpotential is that the solutions $\a=\a^{*}$ of the
equations
\be\label{qem}\frac{\partial\wmic}{\partial
a_{i}}\bigl(\a=\a^{*}\bigr) = 0\ee
are in one-to-one correspondence with the full set of quantum vacua of
the theory \cite{mic1}. The physical correlators in a given vacuum are
then obtained by plugging the corresponding solution $\a=\a^{*}$ to
\eqref{qem} into the generating functions \eqref{Rz}, \eqref{Sz} and
\eqref{Gz},
\begin{align}\label{onshell1} \mathscr R(z) & = R(z;\a^{*})\, ,\\
\label{onshell2}
\mathscr S(z) & = S(z;\a^{*})\, ,\\
\label{onshell3} \mathscr G^{\tilde f}_{\ f}(z) &= G^{\tilde f}_{\
f}(z;\a^{*})\, .
\end{align}
The superpotential $\wmic$ is always unambiguously determined in terms
of the expectation values by a $\u_{\text R}$ symmetry of the model.
In our case, the charges for the relevant $\u_{\text R}$ are given by
\be\label{charges}
\begin{array}{ccccccccccccccl}
&\theta & W^{\alpha} & X & Q_{f} & \tilde Q^{\tilde f} &\a &\g & \m
& U_0 & \b & \t & q & \mathcal W & \\ 
\mathrm{U}(1)_{\mathrm R} & 1 & 1 & 0 & 1 & 1 & 0 & 2 & 0 &0 & 0 & 0
& 0 & 2& , \end{array}\ee
where $\theta$ is a superspace coordinate and $\mathcal W$ an
arbitrary superpotential. The Ward identity associated with this
$\u_{\text R}$ reads
\be\label{wardUR} \wmic(\a) = \sum_{k\geq 0}
g_{k}\frac{\partial\wmic}{\partial g_{k}}\,\cdotp\ee
Using the standard supersymmetric Ward identity
\be\label{wardsusy1}\vevab{\Tr X^{k+1}} = (k+1)
\frac{\partial\wmic}{\partial g_{k}}\, \cvp\ee
we thus get the fundamental formula
\be \label{Wmic} W_{\mic}(\a) = \vevab{\Tr W(X)} \ee
relating the quantum superpotential to the correlators of the chiral
operators $\Tr X^{k}$. We shall also need two additional
supersymmetric Ward identities, similar to \eqref{wardsusy1}, that
read
\begin{align} \label{vmic}-\frac{1}{16\pi^2} 
\vevab{\Tr W^{\alpha}W_{\alpha} X^k} &= \frac{N}{k+1} \frac{\partial
W_{\mathrm{mic}}}{\partial t_{k+1}}\, \cvp \\ \label{wmic}
\vevab{{}^T\tilde Q^{\tilde f} X^k Q_f} &= \frac{\partial
W_{\mathrm{mic}}}{\partial m_{k,\tilde{f}}^{f}}\, \cdotp
\end{align}
\section{The shape function and the scalar operators}
\label{scalarsec}
\begin{figure}
\centerline{\includegraphics[width=12cm]{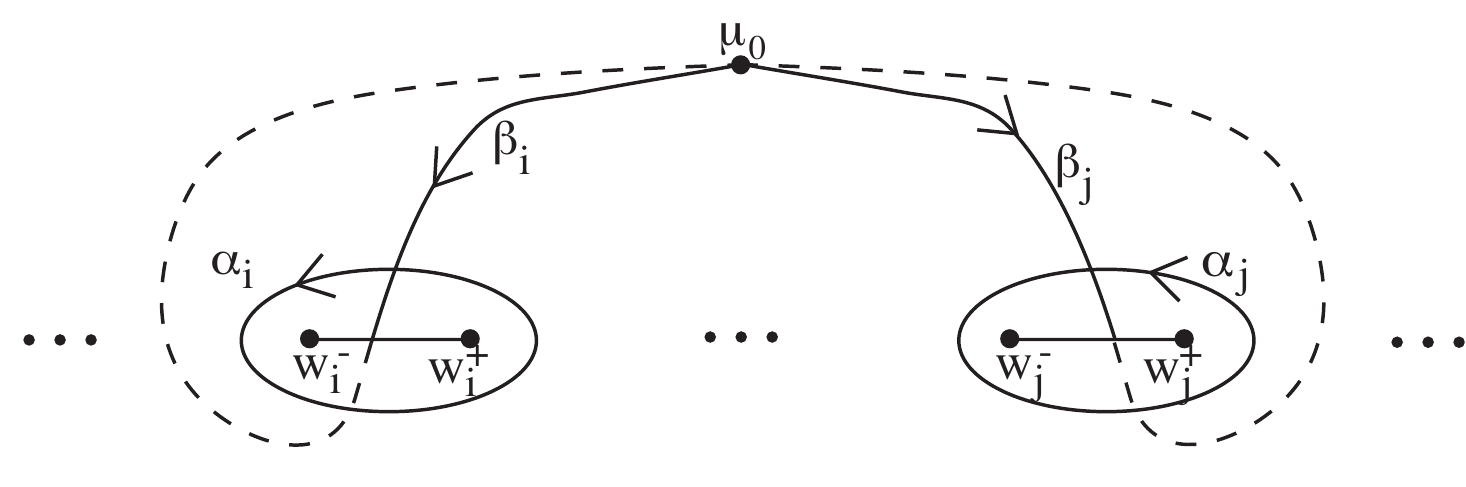}}
\caption{The hyperelliptic Riemann surface $\mathscr C$, with the
cycles $\alpha_{i}$ and chains $\beta_{i}$ used in the main text. Note
that the contours $\alpha_{i}$ are chosen such that they do not
encircle the points $b_{Q}$ (this is needed in Section
\ref{GenMesons}). The point $\mu_0$ is taken to infinity.
\label{fig1}}
\end{figure}

All the chiral correlators for given boundary conditions $\a$ can be
computed using Nekrasov's instanton technology. We plan to provide a
general discussion of this technology in a purely $\nn=1$ context in a
subsequent paper, but for our present purposes all the relevant
formulas can be obtained from the existing literature \cite{nekrasov,
NO,MN} through rather simple generalizations.

It is convenient to present first the solution when all the parameters
in the problem are real, the $a_{i}$s being widely separated and the
$b_{Q}$s sufficiently smaller than all the $a_{i}$s. In this case all
the expectation values can be expressed in terms of a so-called shape
function $f$,
\be \label{Of} \vevab{\mathscr O} = \mathscr O[f]\, .\ee
For example, 
\be \label{Rzf} R(z;\a) =\frac{1}{2} \int_{\mathbb R} \d y \,
\frac{f''(y)}{z-y}\, \cvd \ee
The shape function $f$ extremizes the functional
\begin{multline} \label{energy} \mathcal F = -\frac{1}{8}\int\!
\d x \d y\, (x-y)^2\bigl( \ln |x-y| 
- 3/2 \bigr) f''(x) f''(y) \\ 
+ \frac{1}{4} \int\! \d x\, \sum_{Q=1}^{\ell} (x-b_Q)^2\bigl( 
\ln (x-b_Q) - 3/2 \bigr) f''(x)  
+ \frac{1}{2} \int\! \d x\, t(x) f''(x)\, \end{multline}
with the constraints
\begin{gather}
\label{suppf} \textrm{support}[ f'']  = \bigcup_{i=1}^N\, 
I_i\, , \\ \label{f''}
\int_{I_i}\! \d x \, f''(x) = 2\, , \\ \label{xf''}
\int_{I_i}\! \d x \, xf''(x) = 2a_i\, ,\\
\label{fout} f(x) = \sum_{i=1}^N |x-a_i|\, , \quad \forall x \notin
\bigcup_{i=1}^N\, I_i\, ,
\end{gather}
where the $N$ intervals 
\be \label{Ii} I_i = \left[ w_i^-, w_i^+\right] \ee
are disjoint. The shape function plays a r\^ole that is very similar
to the density of eigenvalues in planar matrix models. In the present
context, the integrals over the instanton moduli space are reduced by
localization to sums over particular field configurations labeled by
colored partitions, and the shape function describes the dominating
colored partition in the limit of vanishing $\Omega$-background
\cite{NO}. A few more details are given in the Appendix \ref{A2sec}.
The correlator $\langle \a | \mathscr O | \a \rangle$ for arbitrary
complex values of the parameters are obtained by analytic
continuation, as will be clear in the following.

The formula \eqref{energy} is a simple generalization of cases that
were previously studied in the literature. The flavorless case of our
model was studied for $\g=0$ in \cite{MN} and for arbitrary $\g$ in
\cite{mic3}. The $\t=\g=0$ theory was studied in \cite{nekrasov, NO}
in the case where the mass matrix $m^{f}_{\ \tilde f}(z)$ is a linear 
function of $z$.

For our purposes, we need to solve the extremization problem in the
general case where all the couplings are turned on simultaneously. The
equation $\delta \mathcal F/ \delta f(x)=0$ reads in this case
\be\label{sp} \int \d y \, \ln |x-y| \, f''(y) = \sum_{Q=1}^{\ell}
\ln (x-b_Q) + t''(x) \, , \quad \forall  x \in \bigcup_{i=1}^N\, I_i\, .
\ee
Instead of working with $f$, it is more convenient to study $R$. Let
us note that \eqref{Rzf} implies that $f$ can be obtained from the
discontinuity of $R$ across the intervals $I_{i}$,
\be \label{disc} R(x+i0)-R(x-i0)=-i\pi f''(x) \, . \ee
The first derivative of \eqref{sp} yields
\begin{align} \label{sdR0} R(x+i0) + R(x-i0) &= \sum_{Q=1}^{\ell}
\frac{1}{x-b_Q} + t'''(x)\\ \label{sdR} & = \frac{U'(x)}{U(x)} +
t'''(x) \, , \quad \forall x \in \bigcup_{i=1}^N\, I_i\, .
\end{align}
This constraint implies that $R$ is a well-defined meromorphic
function on the hyperelliptic curve
\be \label{C} \mathscr C: \
y^2=\prod_{i=1}^N(z-w_i^+)(z-w_i^-) \, . \ee
This curve, with some useful contours, is depicted in Figure
\ref{fig1}. In particular, $R(z)$ is a two-valued function. From now
on, we shall denote by $R(z)$ its value on the first sheet, where the
asymptotic condition at infinity
\be \label{Ras} R(z) = \frac{N}{z}+ \mathcal O\left( 1/ z^2 \right)\ee
is valid, and by $\hat R(z)$ its value on the second sheet. More
generally, a hat on a function defined on \eqref{C} will always mean
that we consider its value on the second sheet. For example, $\hat
y=-y$. Equation \eqref{sdR} is equivalent to
\be \label{RRhat} R(z)+\hat{R}(z) = \frac{U'(z)}{U(z)} + t'''(z) \, .
\ee
As any other meromorphic function on the curve $\mathscr C$, $R$ can
be written in the form
\be \label{RonC} R(z) = r_1(z) + \frac{r_2(z)}{y}\, \cvp \ee
where $r_1$ and $r_2$ are rational functions. Equation \eqref{RRhat}
implies that
\be\label{r1sol} r_{1}(z) = \frac{1}{2}\biggl(
\frac{U'(z)}{U(z)}+t'''(z)\biggr)\, .\ee
Moreover, from the integral representation \eqref{Rzf} and the
constraints on the shape function $f$, $R$ cannot have poles on the
first sheet. From \eqref{RRhat}, we deduce that the only poles of $R$
are on the second sheet at $z=b_{Q}$ with residue one. This implies
that
\be \label{r2P} r_2(z) = \frac{p(z)}{U(z)}\, \cvp \ee
for some polynomial $p(z)$. The asymptotic behaviour \eqref{Ras}
implies that $\deg p = \deg t''' + N + \ell$ and imposes $\deg t''' + 
2$ constraints on the coefficients of $p$. Matching the residues at
the poles on the second sheet implies $\ell$ additional constraints
\be \label{pbQ} p(b_Q) = -\frac{1}{2}U'(b_{Q}) y(b_Q)\, . \ee
There remains $N-1$ unknown coefficients in $p$, as well as the $2N$
unknown branching points $w_{i}^{\pm}$ on the curve \eqref{C}. These
$3N-1$ parameters are fixed by the following $3N-1$ independent
constraints on the periods of $R\,\d z$,
\begin{align} \label{alphaR} \oint_{\alpha_i} R(z)\,\d z &= 2i\pi \, ,
\\ \label{alphazR} \oint_{\alpha_i} zR(z)\, \d z &= 2 i\pi a_i\, ,\\
\label{betaR} \int_{\beta_i} R(z)\,\d z &= \ln\left(
\frac{U_0q}{\mu_0^{2N-\ell}} \right)+NV(\mu_0) + 2i\pi \mathbb Z \, .
\end{align}
The contours $\alpha_i$ and $\beta_i$ are depicted in Figure
\ref{fig1}. The cut-off $\mu_0$ is always understood to be taken to
infinity at the end of the calculations. Let us note that of all the
constraints that determine $R$, none depends on $\g$ and thus $R$
itself will not depend on $\g$ (but will of course depend on $\a$,
$\t$ and $\b$). Note that the physical generating function $\mathscr
R$ will depend non-trivially on $\g$ through the solutions $\a^{*}$ of
\eqref{qem}.

Equations \eqref{alphaR} and \eqref{alphazR} directly follow from
integrating \eqref{disc} and $x$ times \eqref{disc} over the intervals 
$I_i$s and then using \eqref{f''} and \eqref{xf''}. Equation \eqref{betaR}
is more interesting. It comes from the \emph{integrated} form \eqref{sp} 
of the variational equation for the shape function. To see this, let us 
introduce
\be \label{phiz} \phi(z) = \int_{\mu_{0}}^{z} R(z')\,\d z' +
N\ln\mu_{0}\, .\ee
Since we do not specify the contour used to go from the point at
infinity on the first sheet $\mu_{0}$ to $z$ in \eqref{phiz},
\eqref{alphaR} shows that $\phi(z)$ is defined modulo $2i\pi$ on the
first sheet. What happens if we cross one of the branch cuts? Since
$\phi'=R$, we can integrate \eqref{sdR} to find the discontinuity of
$\phi$,
\be\label{phicut}\phi(x+i0) + \phi(x-i0) =
\sum_{Q=1}^{\ell}\ln(x-b_{Q}) + t''(x) + c_{i}\, , \quad\forall x\in
I_{i}\, .\ee
A priori, the integration constants $c_{i}$ could depend on the cut
$I_{i}$. However, by comparing with \eqref{sp}, we find that all the
$c_{i}$s are actually zero. This means that, by crossing any of the
cuts, we go to the same sheet of the function $\phi$. In other words,
modulo $2i\pi$, $\phi$ is well-defined on the curve \eqref{C}. In
particular,
\be \label{phiphihat} \phi(z) + \hat{\phi}(z) = \sum_{Q=1}^{\ell}\ln
(z-b_Q)+t''(z) + 2i\pi \mathbb Z\, . \ee
This yields
\begin{align} \label{phibeta} \int_{\beta_i}R(z)\, \d z &
=\hat\phi(\mu_{0}) - \phi(\mu_{0}) + 2i\pi\mathbb Z\\\label{phibeta2}
&= - (2N-\ell) \ln \mu_0+t''(\mu_0)+ 2i\pi\mathbb Z\, ,
\end{align}
which is equivalent to \eqref{betaR} thanks to the relation
\be \label{t''V} t''(z) = NV(z) + \ln(U_0q)\, , \ee
see \eqref{Vz} and \eqref{tz}.

Equations \eqref{alphaR} and \eqref{betaR} imply that
\be\label{qcoff} \oint R(z)\,\d z \in 2i \pi \mathbb Z\, ,\ee
where the integral is computed along any closed contour on the curve
$\mathscr C$ \eqref{C}. In particular, the function
\be \label{Fphi} F(z) = e^{\phi(z)} \ee
is well-defined on $\mathscr C$. Its value on the second sheet is
determined by \eqref{phiphihat} to be
\be \label{Fhat} \hat{F}(z) =  e^{\hat{\phi}(z)}
=\frac{\prod_{Q=1}^{\ell}(z-b_Q)}{F(z)}\, e^{t''(z)} = 
\frac{q U(z)}{F(z)}\, e^{NV(z)}\, .  \ee

The function $F$ has an essential singularity at infinity on the
second sheet for non-zero $V$. In the special case $V(z)=0$, which
corresponds to the conventional theory with standard gauge kinetic
term, this singularity becomes power-like and $F$ is a meromorphic
function on $\mathscr C$. The solution can then be described more
explicitly. For example, for $\ell\leq 2N$, \eqref{Fhat} implies that
\be \label{FFH} F(z) + \frac{q U(z)}{F(z)} = H_N(z)\, , \ee
where $H_N(z) =(1+qU_0 \delta_{\ell,2N})\, z^N+\ldots$ is a degree $N$
polynomial. Equivalently,
\be \label{Ft0} F(z) = \frac{1}{2}\left( H_N(z) +
\sqrt{H_N(z)^2-4qU(z)}\right)\, . \ee
Comparing with \eqref{C}, we can relate $H_{N}$ to the branching
points $w_{i}^{\pm}$,
\be \label{yY} H_N(z)^2-4qU(z) = \left(1-qU_0
\delta_{\ell,2N}\right)^2\, \prod_{i=1}^N (z-w_i^+)(z-w_i^-)\, . \ee
The generating function $R$ then takes the form,
\be \label{Rt0} R(z) =\frac{F'(z)}{F(z)}=
\frac{1}{2}\frac{U'(z)}{U(z)}+ \frac{1}{\sqrt{H_N(z)^2-4qU(z)}}
\left(H'_N(z)-\frac{U'(z) H_N(z)}{2U(z)}\right)\, \cvd \ee

To finish this Section, let us comment on the analytic structure of
the solution. The structure that we have described above is valid for
generic values of the parameters, but interesting phenomena occur when
the boundary eigenvalues $a_{i}$ of $X$ are chosen to coincide with
the parameters $b_{Q}$. By carefully analysing our solution, it is not
too difficult to show that when $b_{Q}$ approaches $a_{i}$, the cut
$I_{i}=[w_{i}^{-},w_{i}^{+}]$ closes. At $a_{i}=b_{Q}$, the curve
$\mathscr C$ degenerates to a genus $N-2$ curve and the pole at
$z=b_{Q}$ is on the first sheet. More generally, if $p$ distinct
$a_{i}$s are equal to $p$ distinct $b_{Q}$s, the curve degenerates down
to genus $N-1-p$ and the generating function $R$ then has $p$ poles on
the first sheet and $N-p$ poles on the second sheet. We have
illustrated this mechanism on a very simple example in the Appendix
\ref{A1sec}. The cases $a_{i}=b_{Q}$ can actually be treated directly
and most easily at the level of the sums over colored partitions. This
is explained in the Appendix \ref{A2sec}.

\section{The glueball operators}
\label{gluesec}

The inclusion of flavors modifies only slightly the computation of the
generating function $S(z;\a)$ and thus we can follow closely
\cite{mic3}. The fundamental formula relates $S''(z)$ to $R(z)$,
\be \label{SzR} S''(z) = N \sum_{k\geq 1} g_k \frac{\partial
R(z)}{\partial t_k}\, \cvd \ee
This is the same as equation (3.21) in \cite{mic3} and the derivation
given in that reference applies without change when flavors are
included. We can also follow closely \cite{mic3} to derive the
consequences of \eqref{SzR}. The only potential difference in the
analysis could come from the fact that $R$ has poles. However, the
residue of these poles are $\t$-independent and thus they do not enter
in \eqref{SzR}. From \cite{mic3} we thus know that \eqref{SzR} implies
that $S'(z)$ must be a meromorphic function on the curve \eqref{C} of 
the form
\be\label{Sprimeone} S'(z) = \frac{N}{2}W''(z) +
\frac{s(z)}{y}\,\cvp\ee
for a certain polynomial $s$ of degree $\deg W'' + N$. The asymptotic 
condition at infinity on the first sheet
\be\label{asymSprime} S'(z) = \mathcal O\left( 1/z^{2} \right)\ee
yields $\deg W'' + 2$ conditions on $s$. Moreover, one must have
\be\label{S'alpha} \oint_{\alpha_i} S'(z)\,\d z = 0\ee
which yields $N-1$ new independent constraints that determine
completely $s$ and thus $S$.

\section{The generalized meson operators}\label{GenMesons}

We are now going to show that the generating function for the
generalized meson operators \eqref{Gz} is given in terms of the
generating function for the glueball operators that we have just
computed by the formula
\be \label{Gnoint} G^{\tilde f}_{\ f}(z) = \frac{1}{N}\, S(z)\,
\bigl(m(z)^{-1}\bigr)^{\tilde f}_{\ f} - \frac{1}{N}\,
\sum_{Q=1}^{\ell} \frac{S(b_Q)}{z-b_Q}\,
\mathrm{res}_{w=b_Q}\bigl(m^{-1}(w)\bigr)^{\tilde f}_{\ f}\, , \ee
where $m(z)$ is the mass matrix polynomial \eqref{mz}. Note that all
the poles of $G^{\tilde f}_{\ f}(z)$ are on the second sheet.

To do the calculation, it is very convenient to use the variations of
the functional $\mathcal F$ defined in \eqref{energy}. Since $\mathcal
F$ is stationary with respect to the changes of the shape function
$f$, we have the simple formula
\be\label{Fvar} \delta\mathcal F =
\frac{1}{2i\pi}\oint_{\alpha}\delta\psi (z)R(z)\, \d z\, ,\ee
where
\be\label{Deltadef}\psi(z)=
\frac{1}{2}\sum_{Q=1}^{\ell}(z-b_{Q})^{2}\bigl(\ln (z-b_{Q}) -
3/2\bigr) + t(z)\, . \ee
To derive \eqref{Fvar}, we have used \eqref{disc}
and we have defined $\alpha$ to be the sum of the contours that
circle around the branch cuts of the curve \eqref{C},
\be\label{alphadef}\alpha = \sum_{i=1}^{N}\alpha_{i}\, .\ee
In particular, using \eqref{alphazR} and \eqref{Fvar} for $\delta=\partial
/ \partial t_k$, the quantum superpotential \eqref{Wmic} can be
rewritten
\be\label{wmicnewform}\wmic = \frac{1}{2i\pi}\sum_{k\geq
0}\frac{g_{k}}{k+1}\oint_{\alpha} z^{k+1}R(z)\,\d z =
g_{0}\sum_{i=1}^{N}a_{i} + \sum_{k\geq 1}g_{k}\frac{\partial\mathcal
F}{\partial t_{k}}\, \cdotp\ee

We now use the relation \eqref{wmic} combined with \eqref{wmicnewform} 
in the definition \eqref{Gz} to obtain
\be\label{Gder1} G^{\tilde f}_{\ f}(z) = \sum_{k\geq 0}\sum_{k'\geq
1}\frac{g_{k'}}{z^{k+1}}\frac{\partial^{2}\mathcal F}{\partial
t_{k'}\partial m^{f}_{k,\,\tilde f}}\, \cdotp\ee
The partial derivative of $\mathcal F$ with respect to
$m^{f}_{k,\,\tilde f}$ is then evaluated using \eqref{Fvar}. This
yields
\be\label{Gder2} G^{\tilde f}_{\ f}(z) = \sum_{k\geq
0}\sum_{k'\geq 1} \frac{g_{k'}}{z^{k+1}}\frac{\partial}{\partial
t_{k'}}\oint_{\alpha}\frac{\d
w}{2i\pi}\frac{\partial\psi(w)}{\partial m^{f}_{k,\,\tilde f}}
R(w)\, .\ee

The function $\psi$ defined in \eqref{Deltadef} depends on the
$m^{f}_{k,\,\tilde f}$ only through the $b_{Q}$s and $t_{1}$, see
\eqref{defq}. It is clear that $\smash{\partial\psi(w)/\partial
m^{f}_{k,\,\tilde f}}$ does not depend on $t_{k'}$ and thus the partial
derivative with respect to $t_{k'}$ in \eqref{Gder2} acts only on
$R(w)$. We can then rearrange nicely the formula using \eqref{SzR},
\be\label{Gder3} G^{\tilde f}_{\ f}(z) = \frac{1}{N}\sum_{k\geq 0}
\frac{1}{z^{k+1}} \oint_{\alpha}\frac{\d
w}{2i\pi}\frac{\partial\psi(w)}{\partial m^{f}_{k,\,\tilde f}}
S''(w)\, .\ee
The constraint \eqref{S'alpha} ensures that $S(w)$ is single-valued
along the contour $\alpha$. We can thus integrate by part twice in
\eqref{Gder3} to finally get
\be\label{Gder4} G^{\tilde f}_{\ f}(z) = \frac{1}{N}\sum_{k\geq 0}
\frac{1}{z^{k+1}} \oint_{\alpha}\frac{\d
w}{2i\pi}\frac{\partial\psi''(w)}{\partial m^{f}_{k,\,\tilde f}}
S(w)\, .\ee
Now, from \eqref{Deltadef} and using \eqref{Uz} and \eqref{defq} we
get
\be\label{derpsi} \frac{\partial\psi''(w)}{\partial m^{f}_{k,\,\tilde
f}} = \frac{\partial\ln U(w)}{\partial m^{f}_{k,\,\tilde f}} = w^{k}
\bigl(m(w)^{-1}\bigr)^{\tilde f}_{\ f}\, . \ee
The series over $k$ in \eqref{Gder4} can then be summed up easily and 
we find
\be\label{Gder5} G^{\tilde f}_{\ f}(z) =  \frac{1}{2i\pi
N}\oint_{\alpha}\frac{\d w}{z-w} \bigl(m(w)^{-1}\bigr)^{\tilde f}_{\ f}
S(w)\, .\ee
The resulting contour integral can be computed by deforming the
contour $\alpha$ to infinity and picking the contributions from the
poles at $w=z$ and $w=b_{Q}$. This yields the formula \eqref{Gnoint}.

\section{Going on-shell and the anomaly equations}

In this last Section, we are going to solve the equations \eqref{qem}.
Using \eqref{onshell1}, \eqref{onshell2} and \eqref{onshell3}, we
shall then be able to make the link between the off-shell generating
functions \eqref{Rz}, \eqref{Sz} and \eqref{Gz} that we have computed
previously and the physical generating functions \eqref{Rdef},
\eqref{Sdef} and \eqref{Gdef}. In particular, we are going to show
that the latter satisfy the anomaly equations \eqref{a1}--\eqref{a4}.

The starting point is the fundamental formula that relates the
derivative of the quantum superpotential to the $\beta_{i}$ contour
integrals of $S'\d z$,
\be\label{qemder} \frac{\partial\wmic}{\partial a_{i}} =
-\frac{1}{N}\int_{\beta_{i}}S'(z)\,\d z + W'(\mu_{0}) = 0\, .\ee
This relation takes exactly the same form as in the theory with no
flavor, equation (3.51) of \cite{mic3}. The derivation given in this
latter reference, which uses in particular the Riemann bilinear
relations, applies without modification to the present case. This
perfect analogy is due to the fact that the poles, that are a priori
present in the case with flavors, are eliminated when one takes
derivatives with respect to $a_{i}$.

Let us first examine the consequences of \eqref{qemder} for the
glueball operators. For arbitrary values of $\a$, we have seen in
Section \ref{gluesec} that $S'(z;\a)$ was well-defined on the curve
$\mathscr C$ \eqref{C}. Denoting as usual with a hat the value on the 
second sheet, we deduce from \eqref{Sprimeone} that
\be\label{Shat1} S'(z;\a) + \hat S'(z;\a) = NW''(z)\, .\ee
Integrating, we get
\be\label{Shat2} S(z;\a) + \hat S^{(i)}(z;\a) = N\bigl( W'(z) -
W'(\mu_{0})\bigr) + \int_{\beta_{i}} S'(z)\, \d z\, ,\ee
where $\hat S^{(i)}(z;\a)$ denotes the analytic continuation of $S$
through the cut $I_{i}$. For general values of $\a$, $S(z;\a)$ is
\emph{not} defined on $\mathscr C$, since the analytic continuation
through a branch cut depends on the particular branch cut that we
choose. However, for the particular on-shell values $\a=\a^{*}$, the
relation \eqref{qemder} is satisfied and thus the right-hand side of
\eqref{Shat2} no longer depends on $i$. The physical generating
function \eqref{onshell2} is thus well-defined on $\mathscr C$, with
\be\label{Shat3} \mathscr S(z) + \hat {\mathscr S}(z) = N W'(z)\, .\ee
A trivial calculation using this relation immediately implies that the
combination $A(z) = NW'(z)\mathscr S(z) - \mathscr S(z)^{2}$ has no
branch cuts, i.e.\ $\hat A(z) = A(z)$. It cannot have poles from the
discussion of Section \ref{gluesec}. Using the asymptotic condition
$S(z)=\mathcal O(1/z)$ at infinity, we conclude that it must be a
polynomial. This implies the first anomaly equation \eqref{a1}.

We can proceed in exactly the same way to derive the other anomaly
equations. It is straightforward to check that the left hand sides in
\eqref{a2}, \eqref{a3} and \eqref{a4} have no branch cuts by using
\eqref{Shat3}, \eqref{RRhat} and \eqref{Gnoint} (these last two
equations are valid for any $\a$, and thus in particular for
$\a=\a^{*}$). It is also straightforward to check that the residues of
the possible poles all cancel by using the simple pole structure of
the various generating functions that we have discussed in the
previous Sections. The asymptotics at infinity then implies that the
right hand sides of \eqref{a2}, \eqref{a3} and \eqref{a4} must be
polynomials.

Let us close this Section with two remarks. First, we note that the
anomaly polynomials in \eqref{a1} and \eqref{a4} can be obtained by
acting on $\wmic$ with first order differential operators $J_{n}$ and
$L_{n}$ defined exactly as in equations (3.63) and (3.62) of ref.\
\cite{mic3}. In particular, the Riemann bilinear relations used in
\cite{mic3} to make the derivations can be easily generalized to take
into account the poles that appear in the generating functions $R$ and
$G^{\tilde f}_{\ f}$ when flavors are present. On the other hand, to
obtain \eqref{a2} and \eqref{a3} from variations of the microscopic
quantum superpotential, one would have to include arbitrary boundary
conditions for the quarks in the formalism and compute $\wmic$ as a
function of both $\a$ and these quark boundary conditions. The anomaly
polynomials in \eqref{a2} and \eqref{a3} would then follow by acting
on $\wmic$ with suitable first order differential operators containing
partial derivatives with respect to the quark boundary conditions. In
the present paper, we have preferred to integrate out the quarks
exactly first and thus work with a microscopic superpotential that
depends on $\a$ only.

Our second remark concerns the set of solutions to the quantum
equations of motion \eqref{qem}. We have shown that any solution must
satisfy the anomaly equations on top of \eqref{betaR} which is valid
off-shell. Conversely, the set of solutions to the anomaly equations
that also satisfy \eqref{betaR} is known to be in one-to-one
correspondence with the full set of quantum vacua of the theory (see
for example \cite{ferwens,ferphases, CSWphases} and references therein). 
One can
show that all these solutions also automatically solve \eqref{qem},
with one rather trivial exception that is discussed below. A simple
way to understand this point is as follows. First, a straightforward
generalization of the analysis in \cite{mic1} shows that vacua of any
rank $r\geq 1$ of the type $|N_{i},k_{i};\nu_{Q}=0\rangle$ are
automatically included in the set of solutions. Second, one uses the
fact that all the other vacua at the same rank can be obtained by
analytic continuations \cite{ferphases, CSWphases} and thus necessarily 
solve \eqref{qem} as well.

There is an interesting point concerning the vacua having $\nu_{Q}\not
= 0$. At the classical level, one has $\nu_{Q}=1$ when one of the
$a_{i}$ is equal to $b_{Q}$. At the quantum level, one might expect
that the solutions to \eqref{qem} associated with these vacua
correspond also to having $a_{i}=b_{Q}$. This would be natural from
the analysis in the Appendix, that shows that if one imposes the
boundary condition $a_{i}=b_{Q}$, then the quantum function $R(z;\a)$
has a pole at $z=b_{Q}$ on the first sheet. However, what really
happens depends on the cases one considers and can be more subtle. The
subtlety comes from the fact that the variables $\a$ can undergo
non-trivial monodromies, as is well-known from the study of the moduli
space in the $\nn=2$ supersymmetric theories \cite{SW}. Due to these
monodromies, the actual solution $\a=\a^{*}$ to \eqref{qem}
corresponding to a vacuum with $\nu_{Q}=1$ can actually have all the
$a_{i}$s different from the $b_{Q}$s.

The above discussion doesn't apply for the vacua of rank zero. These
vacua have a completely broken gauge group and correspond to the cases
where all the eigenvalues $a_{i}$s are equal to the $b_{Q}$s
classically. From the discussion in the Appendix, we know that the
solution is trivial in these cases: the chiral operator expectation
values do not get any quantum correction. Now, it turns out that these
trivial solutions do not satisfy \eqref{qem}. The reason is that the
procedure of integrating out the quarks become singular from the point
of view of the microscopic quantum superpotential in these particular
vacua. This can be easily illustrated since these vacua are purely
classical. Integrating out the quarks from the tree-level
superpotential
\begin{equation}\label{Wtree}
W_{\text{tree}} = \Tr W(X) + {}^T\tilde Q^{\tilde f} m_{\
\tilde f}^{f}(X) Q_f\ee
amounts to imposing the conditions
\be\label{condquar}
m^{f}_{\ \tilde f}(X)Q_{f} = 0 = {}^T\tilde Q^{\tilde f} m_{\ \tilde
f}^{f}(X)\, .\ee
The resulting effective superpotential, obtained by plugging
\eqref{condquar} into \eqref{Wtree}, is simply
\be\label{wmicclassical}\wmic = \Tr W(X) = \sum_{i=1}^{N} W(a_{i})\,
,\ee 
whose variations only yield $W'(a_{i})=0$. For these solutions, the
matrix $\smash{m^{f}_{\ \tilde f}(X)}$ is invertible and $Q_{f}=\tilde
Q^{\tilde f} = 0$. The superpotential \eqref{wmicclassical} is thus
missing the solutions for which $\smash{m^{f}_{\ \tilde f}}$ has zero
eigenvalues and ${}^T\tilde Q^{\tilde f}Q_{f}\not = 0$. These solutions
correspond precisely to the cases $a_{i}=b_{Q}$. When the rank of the
solutions is $r\geq 1$, and contrary to the case $r=0$, there are
non-trivial quantum corrections and as we have explained above the
solutions are actually obtained from \eqref{qem}.

\subsection*{Acknowledgements}

This work is supported in part by the belgian Fonds de la Recherche
Fondamentale Collective (grant 2.4655.07), the belgian Institut
Interuniversitaire des Sciences Nucl\'eaires (grant 4.4505.86) and the
Interuniversity Attraction Poles Programme (Belgian Science Policy).
Vincent Wens is a junior researcher (Aspirant) at the belgian Fonds
National de la Recherche Scientifique. Frank Ferrari is on leave of
absence from the Centre National de la Recherche Scientifique,
Laboratoire de Physique Th\'eorique de l'\'Ecole Normale Sup\'erieure,
Paris, France.

\renewcommand{\thesection}{\Alph{section}}
\renewcommand{\thesubsection}{A.\arabic{subsection}}
\renewcommand{\theequation}{A.\arabic{equation}}
\setcounter{section}{0} \setcounter{subsection}{0}
\setcounter{equation}{0}

\section*{Appendix: the special cases with $a_{i}=b_{Q}$}

%
\subsection{A simple example: $N=\Nf=1$}
\label{A1sec}

Let us consider the solution for the generating function $R(z;\a)$
described in Section \ref{scalarsec} in the case $N=\Nf=1$, $U=z-b$ and 
$V=0$. If we note $a_{1}=a$ and $b_{1}=b$, we get
\be \label{Rex2} R(z) = \frac{1}{2(z-b)} + 
\frac{1}{\sqrt{(z-a+q)^2-4q(z-b)}} \left(1-\frac{z-a+q}{2(z-b)}\right)
\ee
from \eqref{Rt0} and \eqref{alphazR}. The function $R(z)$ is
two-valued, with asymptotics at infinity $R(z)\sim 1/z$ on the first
sheet and a pole at $z=b$ on the second sheet. The branching points are 
given by the equation
\be\label{bpoints} (z-a+q)^2-4q(z-b) = (z-w^{-})(z-w^{+})\, ,\ee
which yields
\be \label{branch} w^{\pm} = a+q\pm 2\sqrt{q(a-b)}\, . \ee
When $b\rightarrow a$, we see that the two branching points collide and
$R$ reduces to
\be \label{Rex3} R(z) = \frac{1}{z-b}\, \cvd \ee
As explained in the main text, this is a very general phenomenon: when
$b_{Q}\rightarrow a_{i}$, the cut $[w_{i}^{-},w_{i}^{+}]$ closes and
the pole at $z=b_{Q}$ is on the first sheet. We are going to find this
property again in the next subsection from a direct analysis of the
sum over colored partitions.

\subsection{Direct analysis using the sums over colored partitions}
\label{A2sec}

Chiral correlators with fixed boundary conditions $\a$ for the field
$X$ are given by a sum over colored partitions $\cpart$ of the form
\be \label{Ocorr} \vevab{\mathscr O}= \lim_{\eps \rightarrow 0}
\frac{\sum_{\cpart}\nu_{\cpart}\, \mathscr O_{\vec{\mathsf
k}}}{\sum_{\cpart}\nu_{\cpart}}\,\cvp\ee
where the limit $\epsilon\rightarrow 0$ corresponds to a vanishing
$\Omega$-background. A colored partition is a collection
$\cpart=(\k_{1},\ldots,\k_{N})$ of $N$ ordinary partitions
$\k_{i}=\{k_{i,\alpha}\}$, $k_{i,1}\geq k_{i,2}\geq\cdots\geq
k_{i,\tilde k_{i,1}}>0$. Much more details about these sums can be
found for example in \cite{mic2}. The measure over the set of colored
partitions decomposes as
\be \label{nu2} \nu_{\cpart} = q^{|\cpart|}
\mu_{\cpart}^2(\a,\t,\epsilon)
\mathscr E_{\cpart}(\a,\epsilon)\, ,\ee
where the dressing factor $\mathscr E_{\cpart}$ gives the contribution
from the integration over the quark fields,
\be \label{Ek} \mathscr E_{\cpart} = \prod_{i=1}^N
\prod_{\alpha=1}^{\tilde{k}_{i,1}}\prod_{\beta=1}^{k_{i,\alpha}}
U\bigl( a_i + \eps(\beta-\alpha)\bigr)\, . \ee
This formula generalizes the dressing factor obtained in \cite{nekrasov} 
in the case of a linear mass function $m^f_{\ \tilde f}(z)$.
The polynomial $U$ is defined in \eqref{Uz}. When $\epsilon\rightarrow
0$, the sums \eqref{Ocorr} are dominated by a single large colored
partition described by the shape function $f$ \cite{NO}. This has been
used extensively in Section \ref{scalarsec}.

The cases where some of the $a_{i}$s are equal to the $b_{Q}$s are
special. What happens is very clear from the form of the dressing
factor \eqref{Ek}: if $a_{i}=b_{Q}$, only the trivial partition
$\k_{i}=\emptyset$ yields a non-zero contribution. In
general, if $p$ distinct $a_{i}$s are equal to $p$ distinct $b_{Q}$s,
the sum over colored partitions reduces to a sum over $N-p$ ordinary
partitions which can be computed as in Section \ref{scalarsec}. The
dominant colored partition is described by a smooth shape function $f$
that extremizes a functional given by
\begin{multline} \label{energyNu} \mathcal F = -\frac{1}{8}\int
\d x\d y\, (x-y)^2\bigl( \ln |x-y| 
- 3/2 \bigr) f''(x) f''(y) + \frac{1}{2} \int \d x\, t(x) f''(x)
\\ 
+ \sum_{Q=1}^{\ell}\frac{1-2\nu_Q}{4} \int \d x\, (x-b_Q)^2
\bigl( \ln (x-b_Q) - 3/2 \bigr) f''(x)  \, , \end{multline}
where $\nu_Q=1$ if $b_Q=a_i$ and $\nu_Q=0$ otherwise.
The constraints on $f$ are
similar to \eqref{suppf}--\eqref{fout}, except that now the support of
$f$ is made of $N-p$ distinct intervals corresponding to the $a_{i}$s
that are distinct from the $b_{Q}$s. One can solve this
extremization problem as in Section \ref{scalarsec}. The resulting
generating function 
\be \label{RzNu} R(z) = \sum_{Q=1}^{\ell} \frac{\nu_Q}{z-b_Q} + 
\frac{1}{2} \int_{\mathbb R}\d y\, \frac{f''(y)}{z-y} \ee
is defined on a hyperelliptic curve of
genus $N-1-p$, with $p$ poles on the first sheet and $N-p$ poles on
the second sheet having residue one. As already emphasized, this is
exactly the same solution as the one obtained starting from $R(z;\a)$
for generic values of $\a$ and $\b$ and then going to the special
points $a_{i}=b_{Q}$.


%


\begin{thebibliography}{99}
\bibitem{mic1}{F.~Ferrari, \jhep{10}{2007}{065}, arXiv:0707.3885
[hep-th].}
%
\bibitem{mic2}{F.~Ferrari, S.~Kuperstein and V.~Wens,
\jhep{10}{2007}{101}, arXiv:0708.1410 [hep-th].}
%
\bibitem{mic3}{F.~Ferrari, \jhep{11}{2007}{001}, arXiv:0709.0472
[hep-th].}
%
\bibitem{nekrasov}{N.~Nekrasov, \atmp{7}{2004}{831}, hep-th/0206161,
\\ N.~Nekrasov, {\it Seiberg-Witten Prepotential from Instanton
Counting,} Proceedings\ of the International Congress of
Mathematicians (ICM 2002), hep-th/0306211.}
%
\bibitem{CDSW}{F.~Cachazo, M.R.~Douglas, N.~Seiberg and E.~Witten, 
\jhep{12}{2002}{071}, hep-th/0211170.}
%
\bibitem{seifla}{N.~Seiberg, \jhep{01}{2003}{061}, hep-th/0212225.}
%
\bibitem{DV}{
R.~Dijkgraaf and C.~Vafa, {\it A perturbative window into 
non-perturbative physics,} hep-th/0208048.}
%
\bibitem{argfla}{R.~Argurio, V.L.~Campos, G.~Ferretti, R.~Heise, 
\prd{67}{2003}{065}, hep-th/0210291.}
%
\bibitem{per}{F.~Ferrari, \npb{770}{2007}{371}, hep-th/0701220.}
%
\bibitem{ferwens}{F.~Ferrari and V.~Wens, \npb{798}{2007}{470},
arXiv:0710.2978 [hep-th].}
%
\bibitem{ferphases}{F.~Ferrari, \jhep{01}{2009}{026}, arXiv:0810.0816.}
%
\bibitem{CSWphases}{F.~Cachazo, N.~Seiberg and E.~Witten, 
\jhep{04}{2003}{18}, hep-th/0303207.}
%
\bibitem{NO}{N.~Nekrasov and A.~Okounkov, {\it Seiberg-Witten theory and 
random partitions,} hep-th/0306238.}
%
\bibitem{MN}{A.~Marshakov and N.~Nekrasov, \jhep{01}{2007}{104},
hep-th/0612019.}
%
\bibitem{SW}{N.~Seiberg and E.~Witten, \npb{426}{1994}{19}, erratum 
{\bf B 430} (1994) 485, hep-th/9407087; \npb{431}{1994}{484}, 
hep-th/9408099.}
%
\end{thebibliography}
\end{document}